\documentclass{emulateapj}
\usepackage{natbib}
\usepackage{xspace}

\newcommand\ce{Cepheid\xspace}

\shorttitle{Cepheid Templates}
\shortauthors{Yoachim et al.}

\begin{document}
\submitted{{\sc Accepted to AJ:} March 16, 2009}

\title{A Panoply of Cepheid Light Curve Templates}

\author{ Peter Yoachim\altaffilmark{1}, Les P. McCommas\altaffilmark{2}, Julianne J. Dalcanton\altaffilmark{2}, \& Benjamin F. Williams\altaffilmark{2} }

\altaffiltext{1}{Department of Astronomy and McDonald Observatory, University of Texas, Austin, TX 78712; {yoachim@astro.as.utexas.edu}}
\altaffiltext{2}{Department of Astronomy, University of Washington, Box 351580,
Seattle WA, 98195}

\keywords{Cepheids --- distance scale --- Cepheids--- stars}

\begin{abstract}
We have generated accurate $V$ and $I$ template light curves using a
combination of Fourier decomposition and principal component analysis
for a large sample of Cepheid light curves.  Unlike previous studies,
we include short period Cepheids and stars pulsating in the first
overtone mode in our analysis.  Extensive Monte Carlo simulations show
that our templates can be used to precisely measure Cepheid magnitudes
and periods, even in cases where there are few observational epochs.
These templates are ideal for characterizing serendipitously
discovered Cepheids and can be used in conjunction with surveys such
as Pan-Starrs and LSST where the observational sampling may not be
optimized for Cepheids.

\end{abstract}

\section{Introduction}\label{intro}

The Cepheid Period-Luminosity (PL) relation is a fundamental rung in
the astronomical distance ladder.  With the Hubble Space Telescope
(HST) routinely resolving stellar populations in nearby galaxies,
Cepheid distances can potentially be calculated for a large number of
new systems.  In this paper, we show that high quality template light
curves can accurately fit pulsation periods and mean luminosities to
sparsely sampled Cepheid observations.  It is our hope that these
templates can be used to characterize Cepheids in archival as well as
new observations, even when the time sampling of observations has not
been optimized for Cepheid studies.

Principal component analysis (PCA) has found application in a wide
range of astronomical studies \citep[e.g., ][]{Faber73,Conselice06}.
The motivation behind PCA is that it can greatly reduce the number of
variables required to describe a data set.  By performing an
eigenvalue decomposition, PCA can be used to generate a small number
of eigenvectors that describe the majority of the variance in a data set.
In the case of Cepheid variables, one could use PCA analysis to
construct light curves with the smallest possible number of free
parameters.  Ideally, all Cepheids with identical periods would have
a single unique light curve shape.  In this case only a period, an average
magnitude and a phase would be needed as free parameters to fit a
light curve.

The classical technique for determining a stellar pulsation period is
to calculate a ``string length'' \citep{Lafler65,Burke70}.  A light curve is
folded along a trial period and a string length is computed by summing
distances between points consecutive in phase.  The trial period with the
shortest computed string length is taken as the true period.  This method
is accurate for well sampled light curves with small photometric
errors, but has been supplanted by recent techniques.  

More robust periods can be obtained by fitting template light curves.
Fourier analysis of Cepheids was introduced in
\citet{Schaltenbrand71}, and \citet{Stetson96} generated templates
based on Fourier decomposition of MW and LMC stars.  \citet{Tanvir05}
used principal component analysis (PCA) to construct a large set of
well sampled Cepheid observations.  In this paper, we extend the
techniques of \citet{Tanvir05} to include short period Cepheids as
well as first overtone Cepheids.

Short period Cepheids (with periods less than $\sim10$ days) have been
excluded from many studies for a variety of reasons.  First, LMC
Cepheids may show a discontinuity in the PL-relation at 10 days,
casting uncertainty on their utility as reliable standard candles
\citep{Ngeow08}.  Second, there is the possibility of confusion
between fundamental and first-overtone mode Cepheids that have similar
periods but different PL-relations.  Third, if fainter stars like
short period Cepheids are used to estimate distances, incompleteness bias
can skew the derived distances to smaller values
\citep{Sandage88,Freedman01}.  Finally, the shorter period Cepheids
show more variation in light curve shape, making template construction
more daunting.  Keeping these potential problems in mind, we boldly go
forward and derive short period templates regardless.

The outline of the paper is as follows: In Section~\ref{pca_sec}, we
describe generating template light curves using PCA from a large
literature sample of Cepheid stars.  In Section~\ref{err_sec} we
perform Monte Carlo simulations to determine how precisely we can
recover Cepheid parameters using our templates.  In
Section~\ref{dist}, we discuss the process of converting fit
parameters into a distance measure. 

\section{Principal Component Analysis} \label{pca_sec}
\subsection{Training Set Selection}

Following \citet{Tanvir05}, we perform PCA on $V$ and $I$ band light
curves simultaneously.  We therefore gathered light curves for stars
that have both $V$ and $I$ measurements.  The majority of our template
stars come from the Optical Gravitational Lensing Experiment (OGLE)
databases for the LMC \citep{Udalski99a} and SMC \citep{Udalski99b}.
We gathered additional LMC observations from \citet{Sebo02} and
additional LMC and SMC light curves from \citet{Moffett98}.  Our
Galactic Cepheid sample was compiled from light curves in the VizieR
database \citep{Ochsenbein00}, including data presented in
\citet{Berdnikov97} and \citet{Berdnikov01}.  We also included
Galactic Cepheids from the McMaster Cepheid Photometry and Radial
Velocity Data Archive which includes light curves from many sources
\citep{Gieren81,Moffett84,Coulson85,Berdnikov95,Henden96,Barnes97}.
We would have liked to include light curves from the MACHO survey.
However the MACHO survey uses non-traditional ``blue'' and ``red''
filters that are non-trivial to convert to $V$ and $I$.

\begin{figure}
\epsscale{1.2}
\plotone{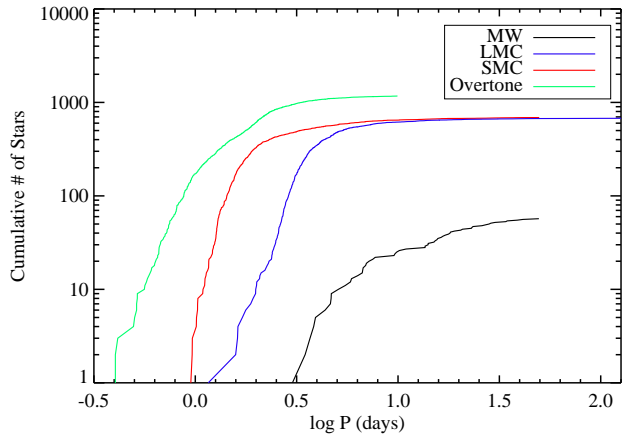}
\caption{Period distribution of the stars that are used to generate our templates.  The Milky Way, LMC, and SMC fundamental mode stars along with the combined LMC and SMC first overtone pulsators are plotted.\label{per_dist}}
\end{figure}

To be included in our analysis we required each light curve have at
least 15 epochs of observations in both $V$ and $I$, with the
exception of well spaced light curves from \citet{Berdnikov01}, where
we only demand 6 epochs in both $V$ and $I$.  The initial sample
contained 3173 light curves, with 248 Milky Way LCs, 1378 LMC LCs, and
1514 SMC LCs.  Unlike the majority of previous Cepheid studies, we did
not exclude stars with periods shorter than 10 days from the analysis.
We fit Fourier components to every light curve and rejected 305 outliers (i.e.,
those with unusual light curve shapes or poor fits).  The final sample
included 150 MW Cepheids 677 LMC Cepheids, 689 SMC Cepheids, and 1171
overtone pulsators from the LMC (549 stars),  SMC (575 stars), and MW (47).
The period distributions of these stars are plotted in
Figure~\ref{per_dist}.

\begin{figure}
\epsscale{1.2}
\plotone{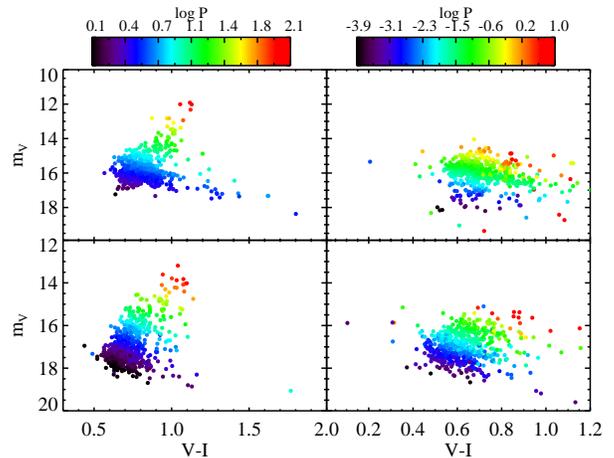}
\caption{Color-magnitude plots for the OGLE stars, uncorrected for reddening. Points have been color-coded by pulsation period. The upper panels show LMC stars while the lower panels show SMC stars.  Fundamental mode pulsators are shown on the left and first-overtones are on the right.  \label{ogle_cmd}}
\end{figure}

\begin{figure*}
\plottwo{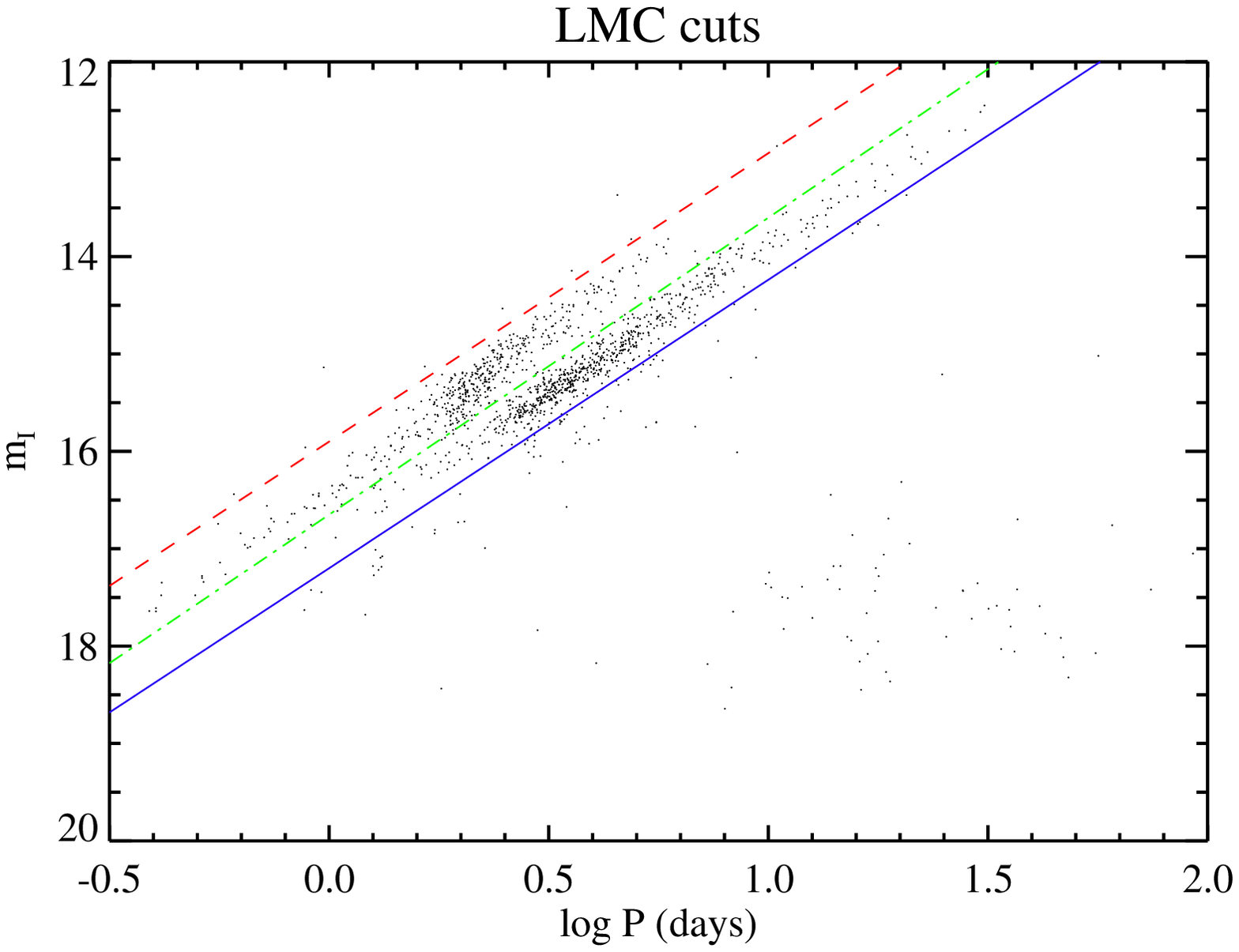}{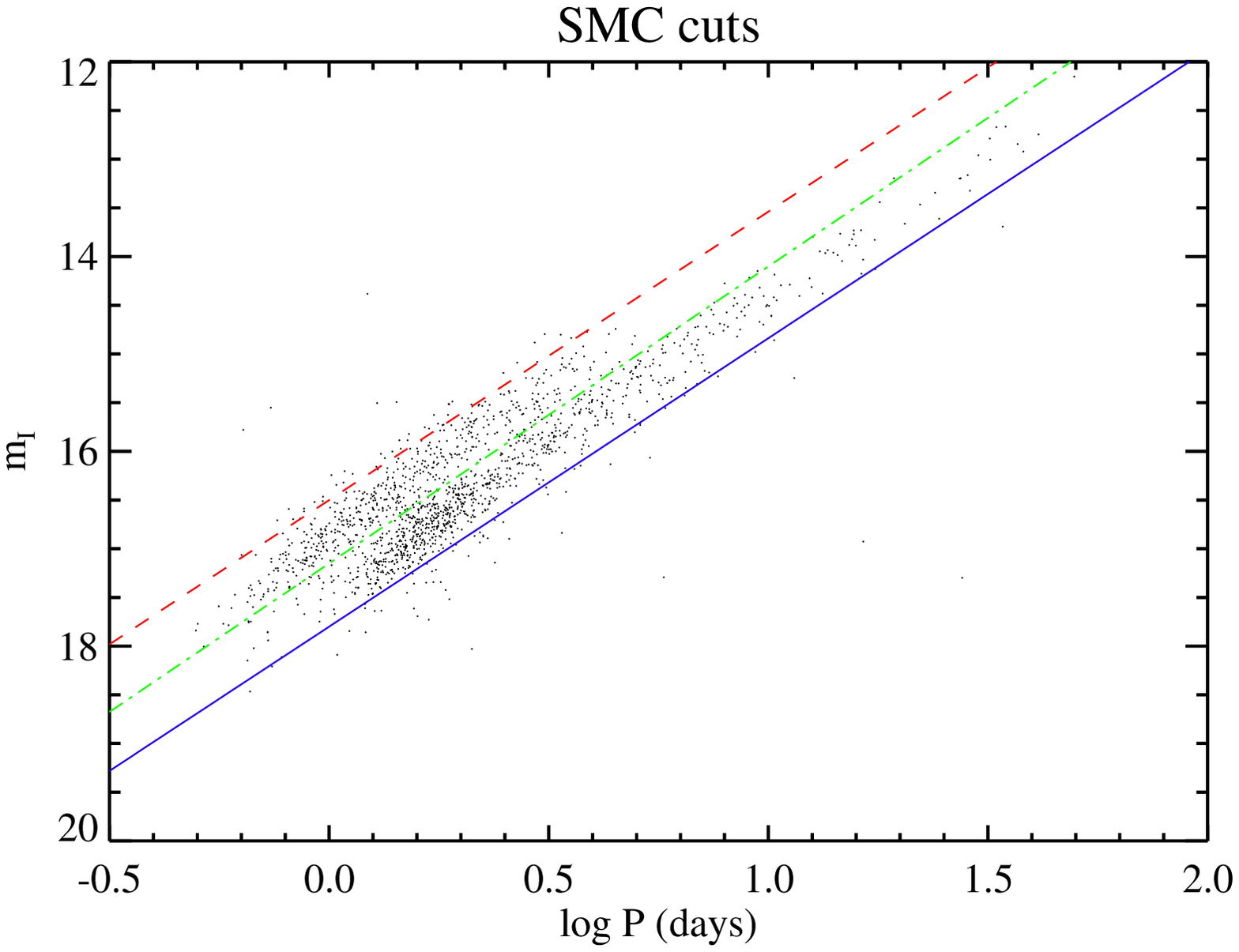}
\caption{First pass classification between fundamental mode and first
overtone from the LMC and SMC OGLE data.  Stars between the solid blue
and dash-dot green line are provisionally labeled as fundamental mode
and stars between the dash-dot green and dashed red are labeled as
first-overtone.  The lines have a slope of -3 and y-intercepts of
17.2,16.65,15.9 for the LMC and 17.8,17.15, and 16.5 for the
SMC.}\label{cuts}
\end{figure*}

The OGLE database does not distinguish between fundamental mode
Cepheids and Cepheids pulsating in the first overtone mode.
Figure~\ref{cuts} shows our simple cuts in period-magnitude space used
to provisionally classify OGLE stars as fundamental or overtone
pulsators.  For the rest of the stars, we use the source catalog
designation of overtone or fundamental Cepheid.  Color-magnitude diagrams of the OGLE are plotted in Figure~\ref{ogle_cmd}.

We initially decompose each light curve in our sample into Fourier
components by fitting the following equations to the $V$ and $I$ light
curves simultaneously.
\begin{eqnarray}
m_I(t)=m_{I,0}+\sum_{k=1}^{k=8}\alpha_k\sin{(2\pi kt/P)}+\beta_k\cos{(2\pi kt/P)}\\
m_V(t)=m_{V,0}+\sum_{j=1}^{j=8}\alpha_j\sin{(2\pi jt/P)}+\beta_j\cos{(2\pi jt/P)}.
\end{eqnarray}
where $P$ is the period, $t$ is the Julian day of the observation,
$m_{I,0}$ and $m_{V,0}$ are the average magnitudes in each band, and
the $\alpha$ and $\beta$ terms are the Fourier amplitudes.  We
constrain $\beta_{j=1}$ to zero to impose a common phase for all the
fits.  These Fourier fits generate the 32 $\alpha$ and $\beta$ values
for each star that are used in the principal component analysis.
While the shapes of the light curves are fitted simultaneously, the
average magnitudes in each band are completely independent.  This
ensures our light curve templates are not dependent on the various
dust corrections (or lack thereof) that have been applied in our
different source catalogs.  

\begin{figure}
\epsscale{1.2}
\plotone{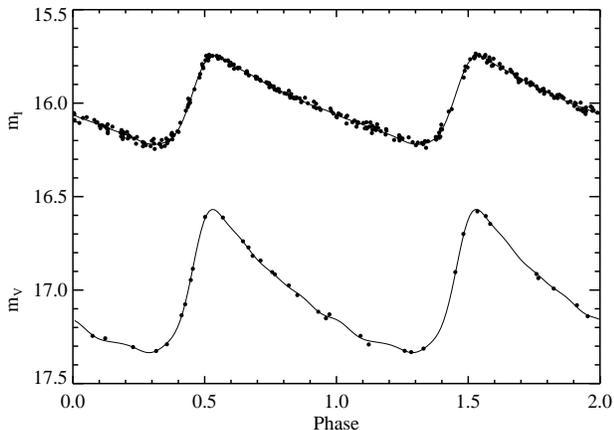}
\caption{Example of an OGLE 2.8 day period \ce with the solid curves
showing the best fit Fourier decomposition.  The upper curve shows the
$I$-band while the lower curve shows $V$.  The sparse sampling in the
$V$-band around a phase of 1.0 and 1.2 allows the fit slightly too
much freedom (The curve rises slightly in those two regions).  By
performing the PCA analysis over many hundreds of light-curves, such
discrepancies should be averaged out. }\label{example1}
\end{figure}

\begin{figure*}
\epsscale{1.2}
\plotone{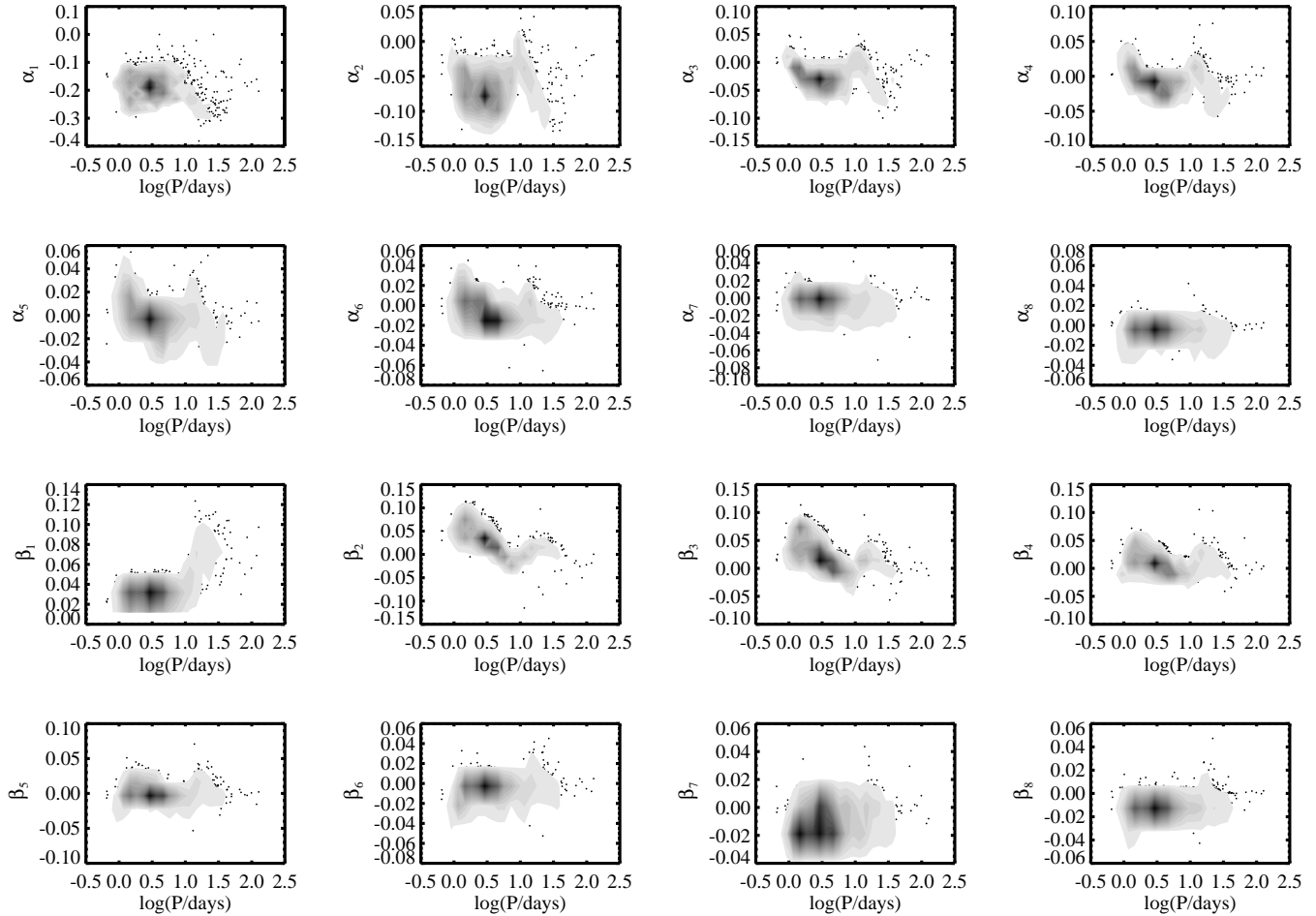}
\caption{The distribution of Fourier decomposition parameters for the best fits to all fundamental mode
Cepheids in the $I$-band.  For each population of stars, we fit low
order polynomials to these Fourier component versus period plots and
reject stars from further analysis that are more than 3$\sigma$
outliers.  The number of stars in our sample is large enough that we
have plotted contours in the densely populated regions and individual points in the sparsely sampled regions.
\label{ifufits}}
\end{figure*}

\begin{figure*}
\epsscale{1.2}
\plotone{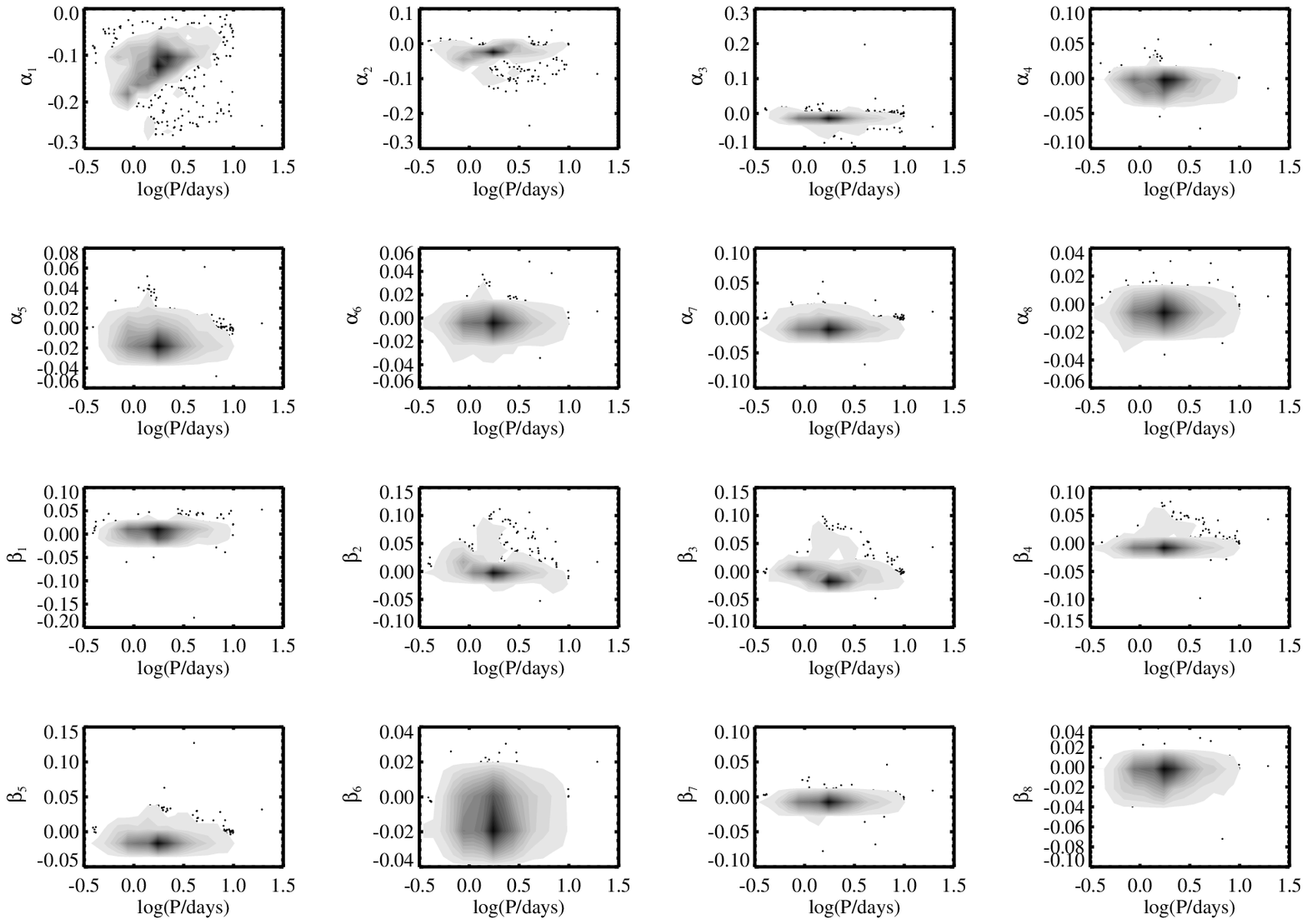}
\caption{Best fit Fourier components for the overtone Cepheids in the $I$-band.  The distributions for the high order coefficients are concentrated near zero, implying these light curves are simply sinusoidal, an unsurprising result for stars commonly referred to as ``sinusoidal'' or sCepheids. \label{ifofits} }
\end{figure*}

\citet{Ngeow03} discuss how fitting Fourier components to sparsely
sampled data can result in poor fits (see their Figure~2b for an
example of how Fourier decomposition can fail for sparsely sampled
light curves).  To keep our fits constrained to reasonable shapes, we
create a smooth light curve by linearly interpolating each observed
Cepheid to include 50 points evenly distributed across the full phase
of the light curve.  We then fit the Fourier components of this
smoothed light curve, and use the results as the initial guess for the
fitting of the observed data points.  In the final fit, each Fourier
component is allowed to change by a maximum of 20\% from the smoothed
fit parameters, thereby preventing divergences compared to the smooth
fit.  Figure~\ref{example1} shows typical results for out fitted
Fourier components.  Figures~\ref{ifufits} and~\ref{ifofits} show the
distribution of the 16 best fit $\alpha$ and $\beta$ values for all of
the fundamental and overtone $I$-band light curves.

\subsection{PCA Template Construction}\label{pca_construct}

\begin{figure*}
\plotone{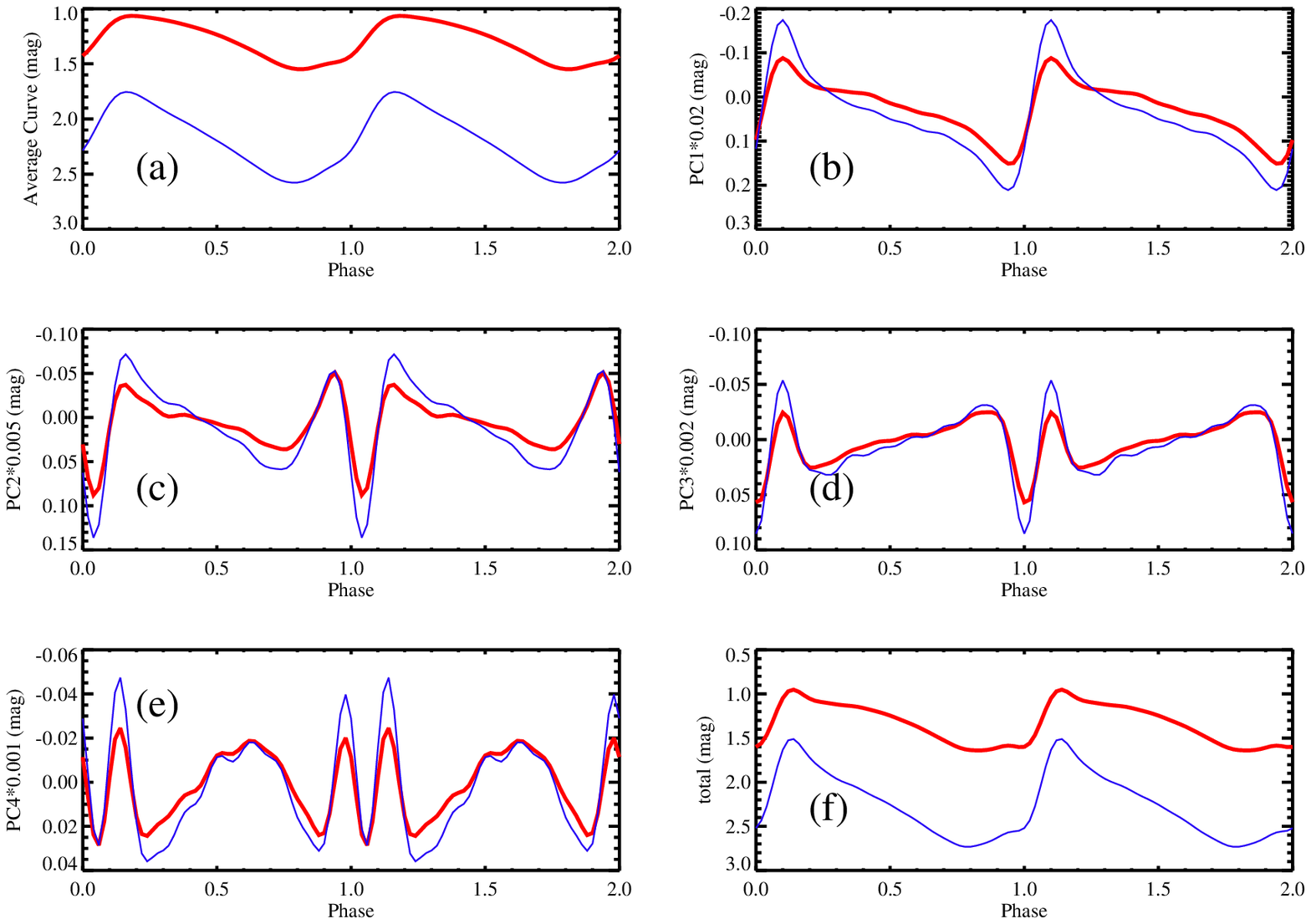}
\caption{ The average light curve (a) and eigenvectors from the
fundamental Cepheid principal component analysis.  $I$-band is shown
in thick red while $V$-band is shown in blue.  The first PCA vector (shown
in b) primarily controls the magnitude of pulsation, while the other
PCA vectors (c-e) control the exact shape of the sawtooth rise of the
light curve.  Template light curves are constructed as linear
combinations of these vectors, an example of which is shown in (f).
Typical coefficients for these vectors are plotted in
Figure~\ref{pcas}.  \label{pca_shapes}}
\end{figure*}

To test for differences in Cepheid populations, we constructed
templates based on various subsets of the data.  In particular, we
made templates that include all our fundamental mode Cepheids and
subsets which included only LMC stars, only SMC stars, only short
period stars, only long period stars, and only overtones.  For each
subsample, we first subtracted an average light curve from each
measured light curve.  We then performed a PCA on the residuals to
develop eigenvectors that can be used to correct for deviations from
the average light curve.  In all cases, over 90\% of the variance
could be described with only four principal component eigenvectors
(Table~\ref{pca_var}).  An example of an average light curve and PCA
eigenvectors are plotted in Figure~\ref{pca_shapes}.

\begin{deluxetable}{ l c c c c c}
\tabletypesize{\small }%\footnotesize \scriptsize}
%\rotate
\tablewidth{0pt}
%\tablenum{num}
%\tablecolumns{num}
%\tableheadfrac{num}
\tablecaption{Percentage of the total variance contained in each PCA component for each of our constructed templates. \label{pca_var}}
\tablehead{ \colhead{Model} &\colhead{PCA1}  & \colhead{PCA2}  
  & \colhead{PCA3}  & \colhead{PCA4} & \colhead{Total}  \\
\colhead{} & \colhead{\%}   & \colhead{\%}  & \colhead{\%}  & \colhead{\%} & \colhead{\%}}
\startdata
short period    &  70.1   &     18.7  &   4.8    &   1.4  & 95.1\\
long period    &   64.1   &     16.6  &   6.8    &   2.9 &  90.4 \\
LMC            &   66.7   &     15.6  &   9.8    &   1.8 &  94.0\\
SMC            &   70.1   &     17.2  &   6.0    &   1.5 &  94.8\\
overtones       &  82.8   &     6.0   &   3.0    &   1.0 &  92.9\\
\enddata
\end{deluxetable}

\begin{figure*}
\epsscale{.325}
\plotone{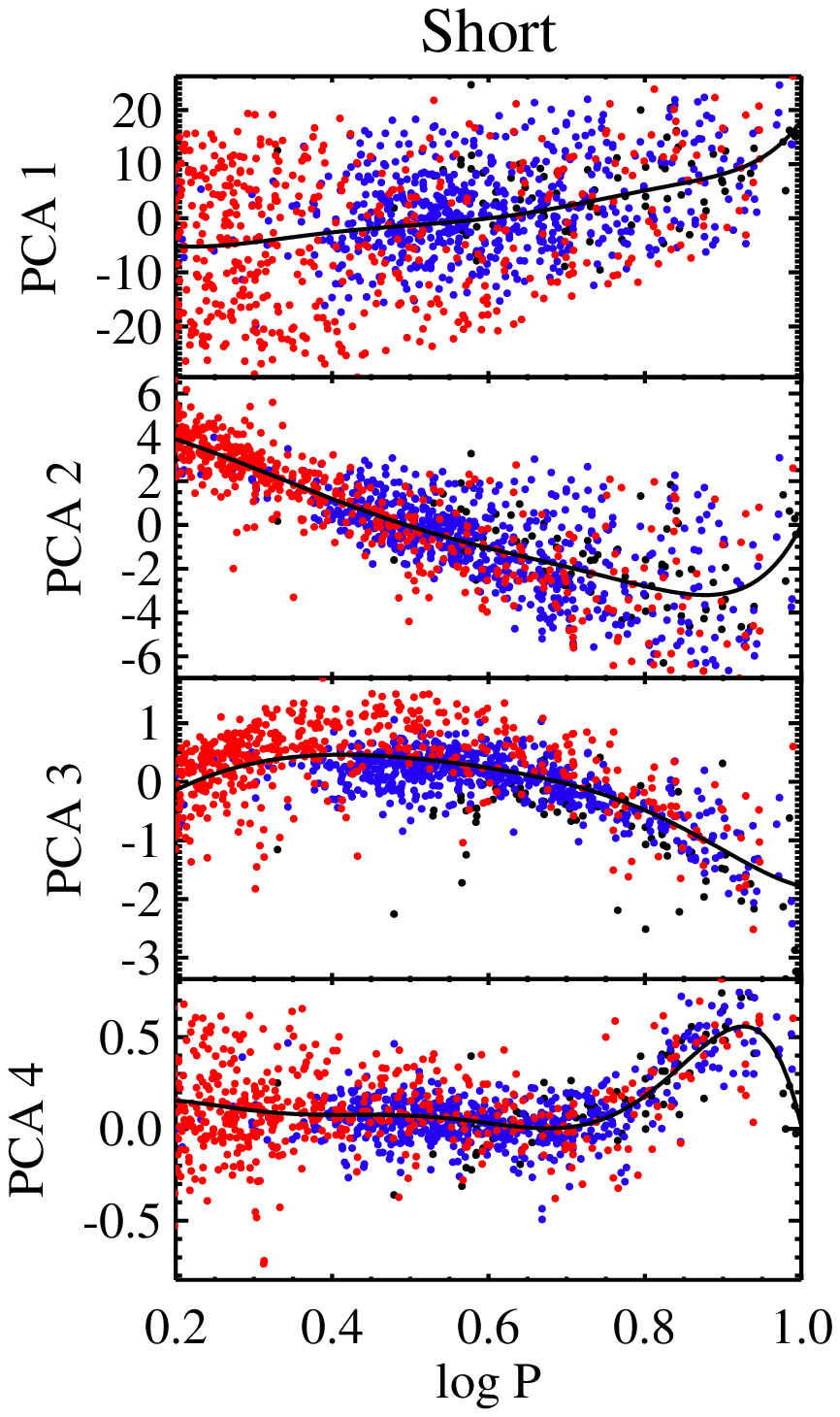}\plotone{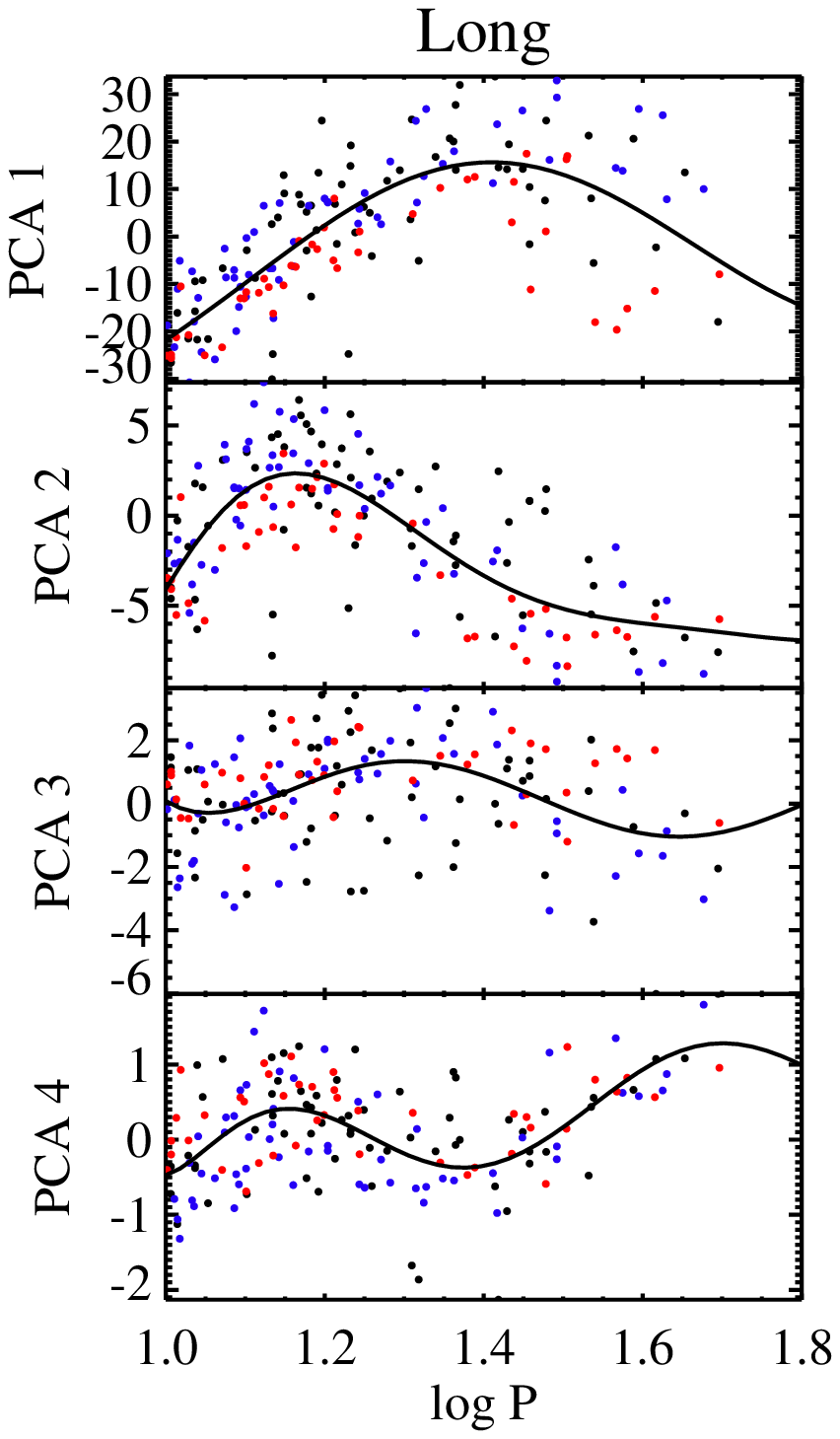}\plotone{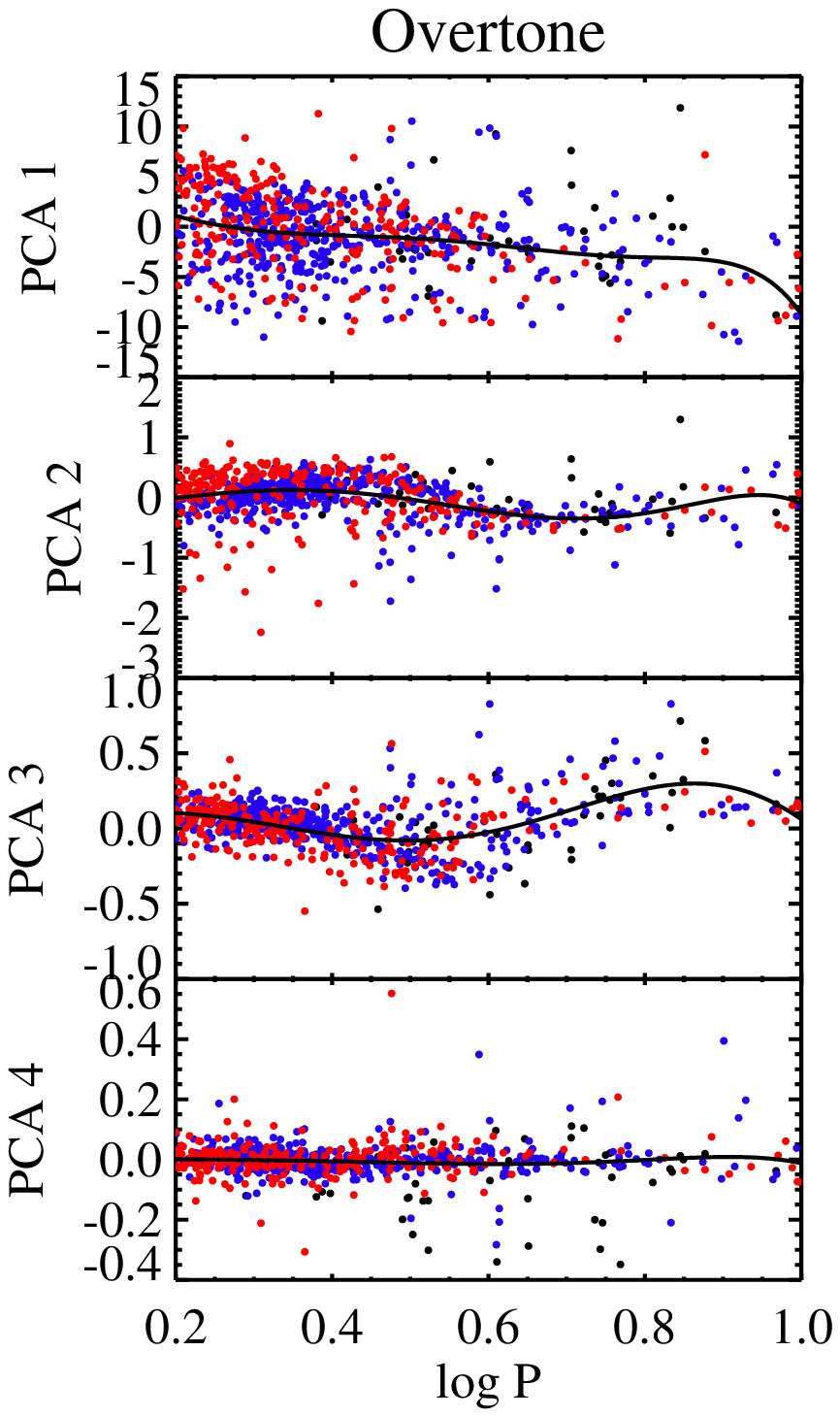}
%\plotone{f8b.eps}
\caption{ The template PCA coefficients (all multiplied be a factor of $10^3$) for all the fundamental mode Cepheids.  The SMC, LMC, and MW stars are shown by red, blue, and black points respectively.  Our polynomial fits are plotted as solid curves.  We have separated long period, short period, and overtone Cepheids into the left, middle, and right panels respectively.\label{pcas}}
\end{figure*}

We plot the PCA vector strengths as a function of period in
Figure~\ref{pcas}.  Several of the PCA vectors show strong trends with
period.  This is fortunate (and not too surprising), as it means while
fitting for a star's period, we can simultaneously make an educated
guess as to what the corresponding shape parameters should be.

Before performing PCA, we needed 32 Fourier components, as well as two
magnitudes, a period and a phase to accurately fit a Cepheid light
curve.  After performing PCA $>$90\% of the variation in the light
curve's shape can be described with just four eigenvector amplitudes.
As a final step we note that the eigenvector amplitudes are strong
functions of period, allowing us to make a quality template fit with
only four free parameters (two average magnitudes, a phase, and a period).
Figure~\ref{pcas} does show one possible pitfall as the first PCA
vector for the short period stars shows a great deal of scatter and
little trend with period.  We therefore caution that it is possible
this eigenvector amplitude should be left as a free parameter if
possible when fitting a light curve.

\begin{figure*}
\epsscale{.85}
\plotone{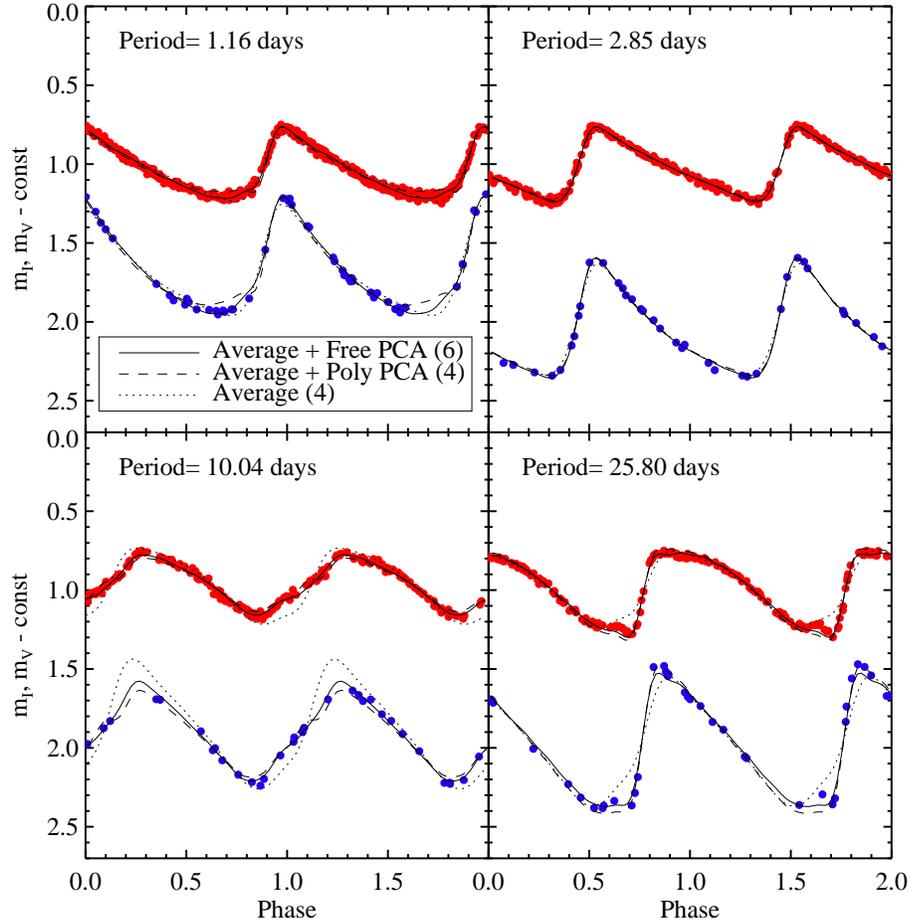}
\caption{ Examples of four OGLE fundamental mode Cepheids fit with our
derived templates.  Both the $V$ and $I$ light curves are fit
simultaneously.  Dotted lines show the best-fitting curve if we fit
with just the average light-curve (PCA eigenvectors fixed at zero),
dashed curves show the best fit if we set the PCA eigenvectors to the
best polynomial fit values for the given period, solid curves show the
fit if we leave the first two PCA vectors as free parameters.  The
numbers in parentheses show the total number of free parameters in
each fit.  }\label{example2}
\end{figure*}

\begin{figure*}
\epsscale{.85}
\plotone{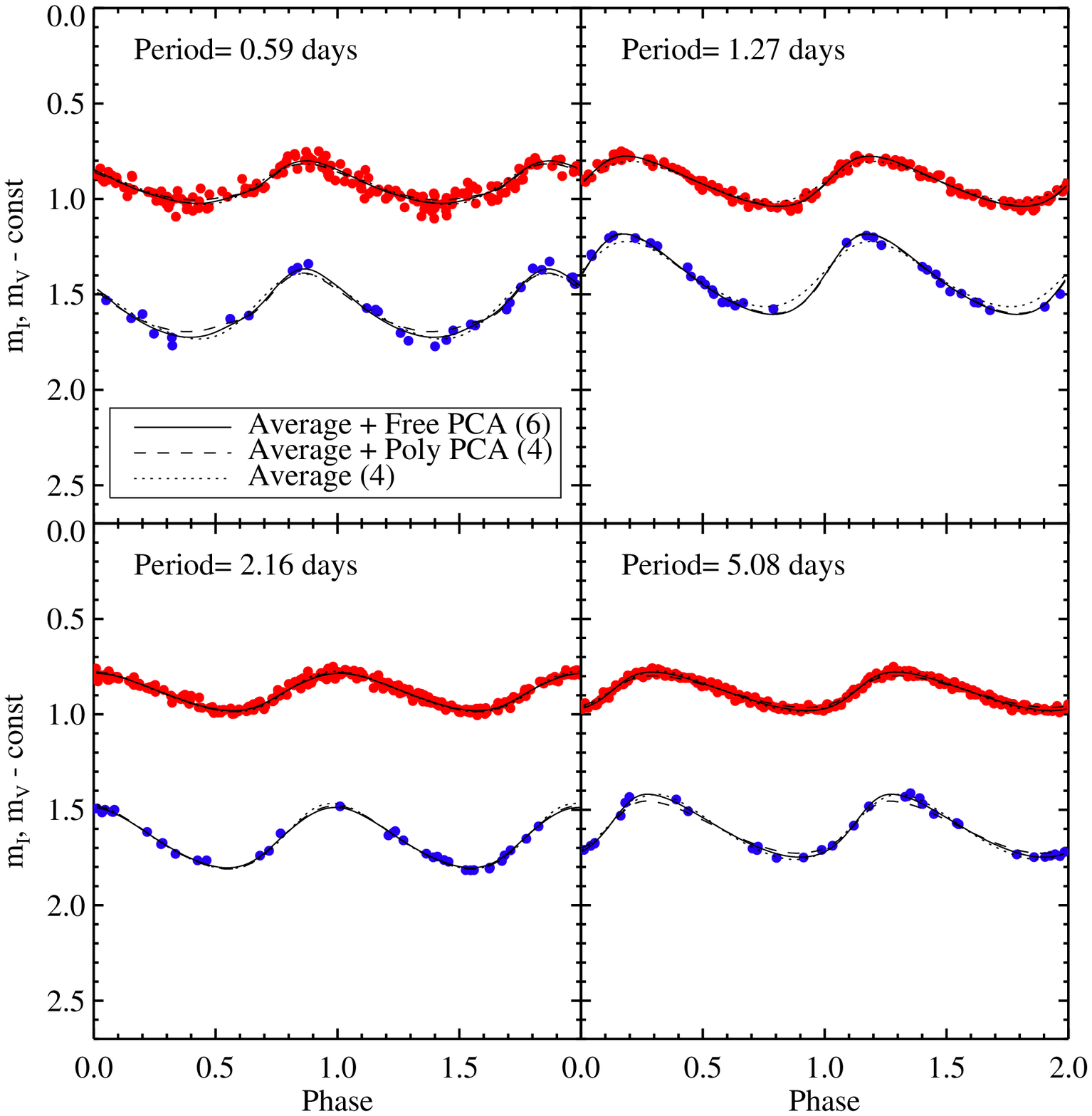}
\caption{Same as Figure~\ref{example2}, only now for Cepheids
identified as first-overtones and fit with our overtone
template. Again, it does not seem necessary to leave the PCA amplitudes as free parameters to converge on a quality fit.}\label{example3}
\end{figure*}

\section{Template Fitting Accuracy}\label{err_sec}

\citet{Tanvir05} have already demonstrated that the template fitting
technique is superior to other common period estimation techniques for
cases where the photometry is noisy. They show magnitudes and periods
determined through template fitting can reduce the scatter in distance
estimates by 30\% compared to simple string length methods.

We now endeavor to use Monte Carlo simulations to quantify how well
our templates can recover magnitudes and periods from photometry of
Cepheid stars.  We have developed a fitting routine that uses
Levenberg-Marquardt least-squares minimization to find the best
fitting Cepheid light curve, given a set of $V$ and $I$ photometry.
The best fitting template is found by varying period, phase, $<I>$,
and $<V>$.  We have included an option to vary the amplitude of the
PCA eigenvectors.  Fitting sparsely sampled light curves, there is a
risk of aliasing or converging on local $\chi^2$ minima.  To avoid
such problems, we use a series of initial guess periods and phases to
ensure we find the global $\chi^2$ minimum.

Our Monte Carlo varies four parameters to judge their impact on our
fitting routine's robustness:  First, we compare 5 well sampled Cepheids
from the OGLE database with different periods;  Second, we vary the total
number of observations in each band;  Next, we look at possible effects
of template mis-match (fitting LMC stars with a template derived from
SMC stars, fitting overtone stars with fundamental mode templates and
vice-versa);  Finally, we vary the photometric precision of the light curve
points.

\begin{deluxetable}{ c c c c}
\tabletypesize{\small }%\footnotesize \scriptsize}
%\rotate
\tablewidth{0pt}
%\tablenum{num}
%\tablecolumns{num}
%\tableheadfrac{num}
\tablecaption{ The different epoch samplings used in our Monte Carlo error analysis.  \label{obs_strat}}
\tablehead{ \colhead{Model} &\colhead{$I$ observations}  & \colhead{$V$ observations} 
  & \colhead{total}}
\startdata
1 & 20 & 15 & 35\\
2 & 10 &  5 & 15\\
3 & 5  &  3  & 8 \\
4 & 4  &  2  & 6\\
\enddata
\end{deluxetable}

For each realization the star was randomly sampled in both bands.
Therefore there are $N_I+N_V$ unique epochs of observations and
we do not explicitly model observing strategies that observe in
multiple filters simultaneously.  

%These simulations also do not
%explicitly address how well template fitting recovers parameters when
%the light curve is well sampled over only a limited portion of the
%full phase.

We used four fundamental mode Cepheids (periods of 2.5, 4.2, 16.0, and
20.7 days), and one overtone (period 2.5 days) from the OGLE LMC data
set for the Monte Carlo tests.  All the photometry has initial errors
of order 0.013 mags in both bands.  Although there could be some
objection to using stars which are included in the PCA analysis, our
sample sizes are large enough that the addition or subtraction of a
few stars should make little difference to our final templates.

For the error analysis, we fit for only period, $m_v$, $m_I$, and
phase.  We rely on the polynomial fits in Figure~\ref{pcas} to give
reasonable amplitudes for the PCA eigenvectors.  We also folded the
original light curves so that the photometry covers at most 3 periods.
This restriction is needed to keep the fitting procedure from falling
into local minima caused by aliasing.  When fitting sparsely sampled
data over a long baseline a more brute-force exploration of period
parameter space would be required than our current fitting procedure.

\begin{figure*}
%\plottwo{f11a.eps}{f11b.eps}
\epsscale{.7}
\plotone{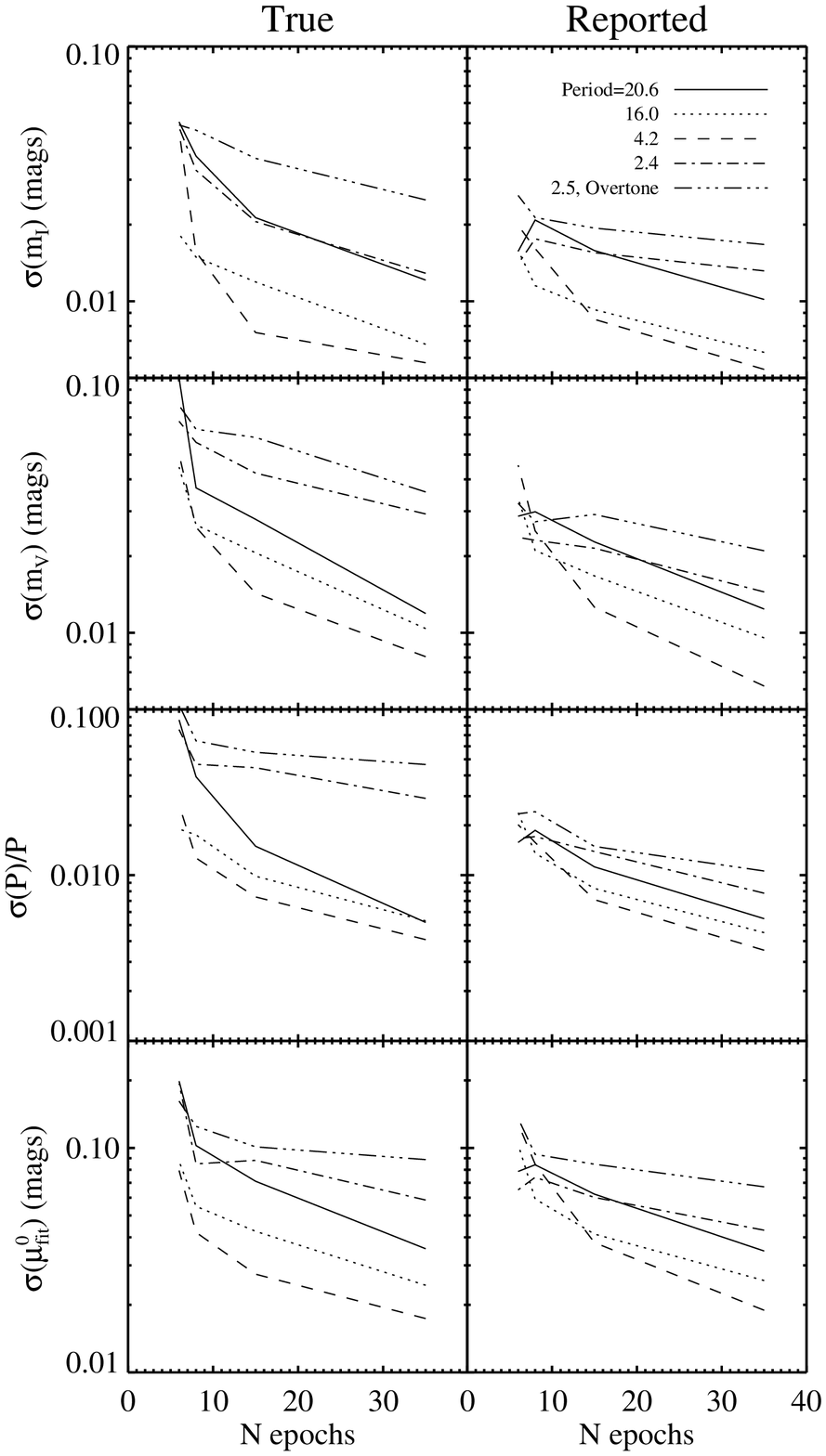}
\caption{ Results of our Monte Carlo simulation where we recover Cepheid properties using our templates and least-squares fitting procedure.  We fit 5 different stars, along with the 5 different observing strategies listed in Table~\ref{obs_strat}.  The four panels on the left show the errors in fitting magnitudes and periods and the corresponding error in distance modulus.  On the right, we show the uncertainties returned by our least-squares fitting routine.  With photometric errors of 0.01 mags and only 6 total epochs of observations, the uncertainties of the fitted parameters result in only a 0.1 mag uncertainty in $\Delta\mu^0_{fit}$.  \label{err_sim1}}
\end{figure*}

Figure~\ref{err_sim1} shows an example of one of our fitting
simulations.  As expected, the errors are largest for the overtone
Cepheid when fit with the wrong template.  We also show the calculated
uncertainties reported by our fitting routine in
Figure~\ref{err_sim1}.  In general, the reported uncertainties are a
good match to the actual errors resulting from the fits.  The one
exception is that the overtone uncertainties are under-estimated as a
result of making the assumption that the reduced $\chi^2$ should be
unity, which is clearly incorrect in the case of fitting an overtone
with a fundamental mode template.  When we use an overtone template, there
is a clear improvement in the $\chi^2$ values, indicating that it is a
better fit.

We have also explored possible template mismatches due to metallicity
by repeating the Monte Carlo experiment using an SMC template
([Fe/H]$\sim-0.7$) to fit LMC stars ([Fe/H]$\sim-0.3$).
The resulting periods and magnitudes are practically identical,
suggesting that the templates can be used across different metallicity
populations.

\begin{figure*}
%\plottwo{f12a.eps}{f12b.eps}
\epsscale{.7}
\plotone{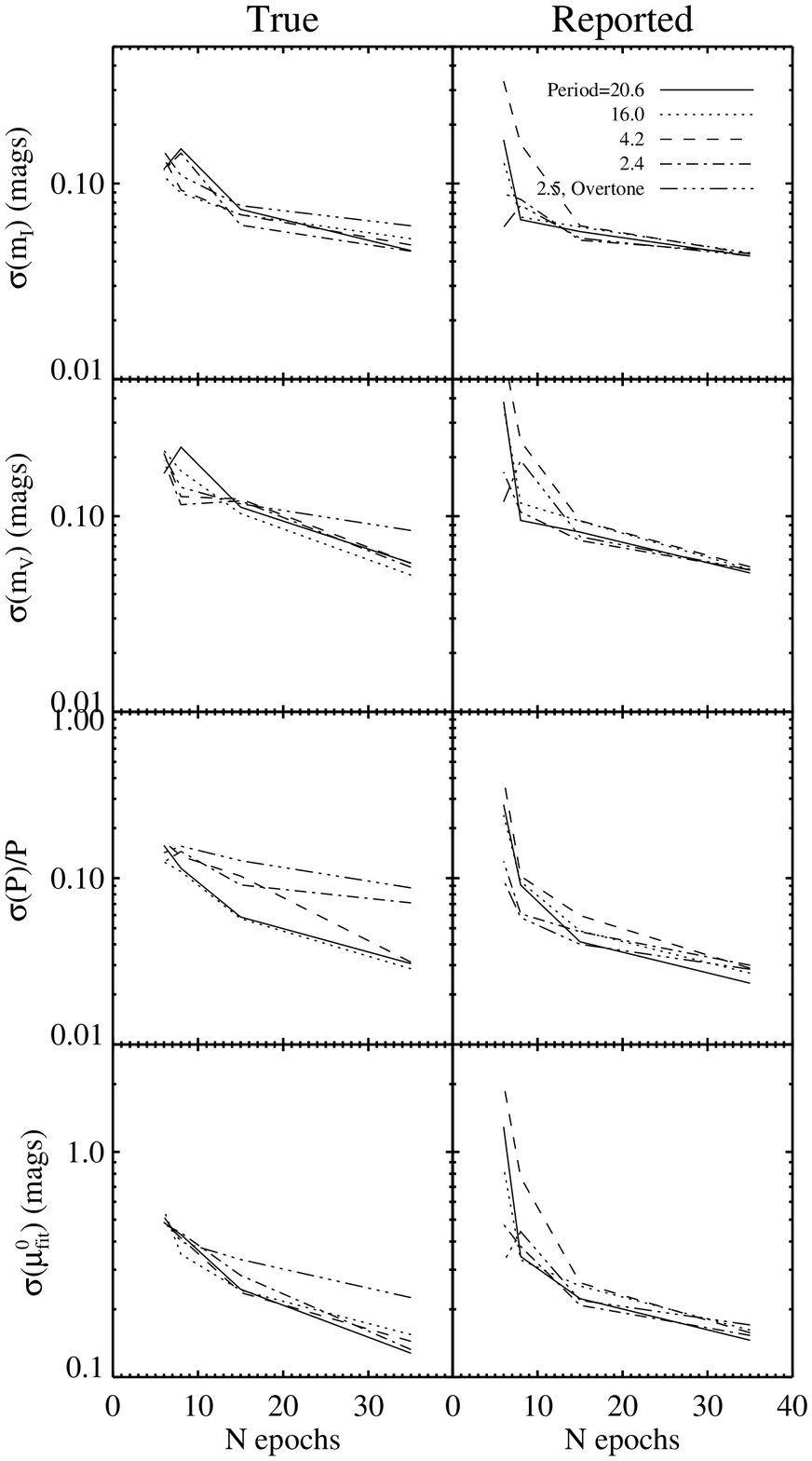}
\caption{Same as Figure~\ref{err_sim1}, only now the photometric errors have been increased to 0.2 mag. Again, the actual errors (left side) and very similar to the returned uncertainties (right side).  \label{err_sim2}}
\end{figure*}

A proper assessment of the errors of individual measurements is key for determining a proper $P$-$L$ relation.  We have therefore compared the uncertainties reported by our fitting procedure to the true offsets seen in our Monte Carlo re-sampling.  Overall, the returned uncertainties are comparable to the actual errors derived from the Monte Carlo analysis.  When the photometric errors were low (0.01 mag), the returned uncertainties were slightly too small (by $\sim$15\%).  On the other hand, when the errors were high and the sampling was sparse, the returned uncertainties were slightly larger than the true Monte Carlo calculated errors.  Typically, around 10\% of the fits would fail catastrophically, usually caused by aliasing or very sparse sampling.  These failures could readily be seen as poor $\chi^2$ values or by visual inspection.  This seems to imply that we can use the returned uncertainties, but as we will see later, observing strategy also affects the results and it is probably best to simply run a Monte Carlo for the photometric error and uncertainties for the specific sampling frequency used by a given observing program.

We ran several simulations to test  how sensitive our fits are to the phase coverage of the observations.  Generally, if the observations do not span more than half of the full phase, there is a large likelihood (10\%-40\%) that the fitted period will catastrophically fail (defined as a final fitted distance error of  $>$10\%).  

\begin{figure*}
%\epsscale{1.2}
\plottwo{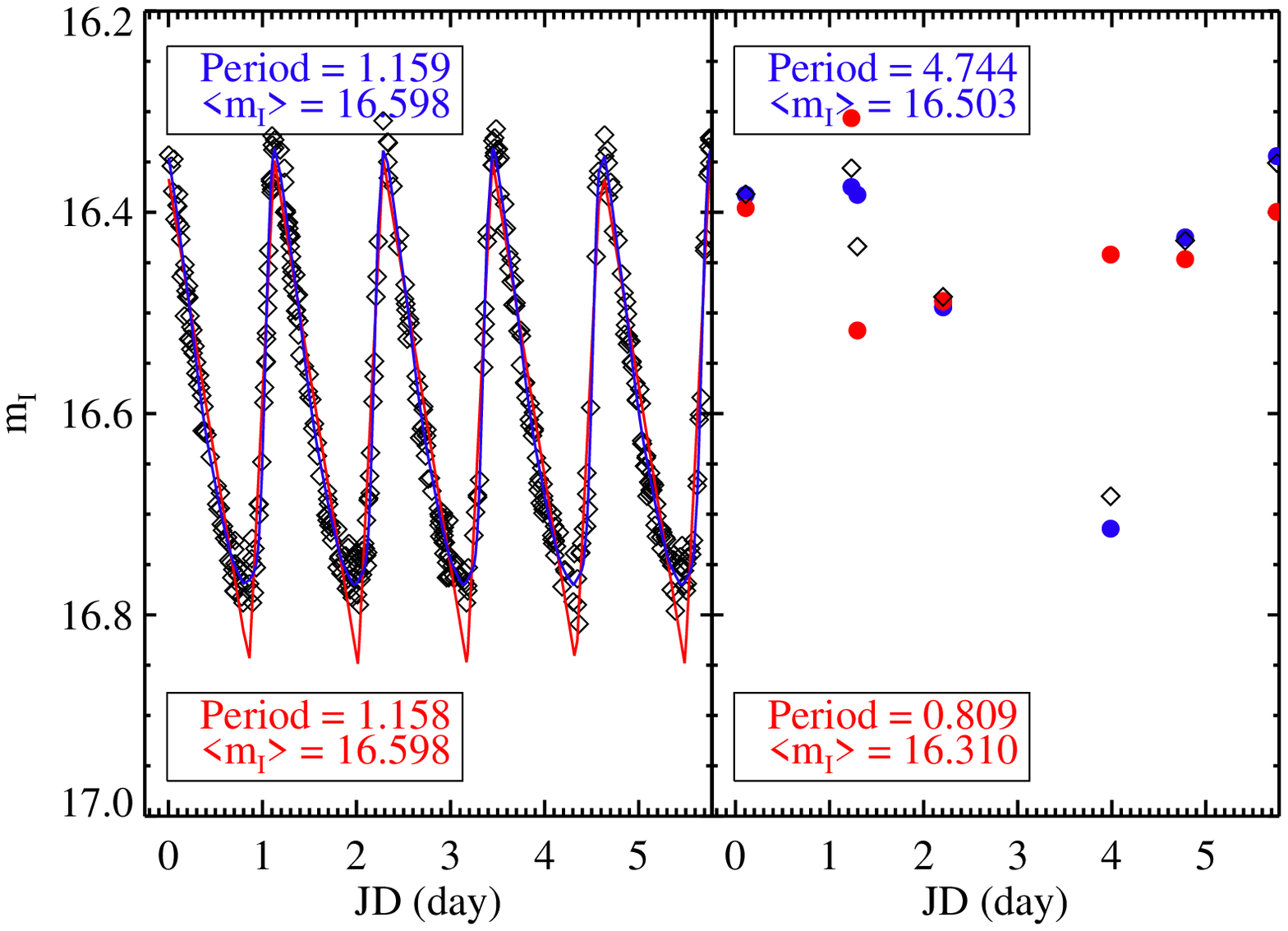}{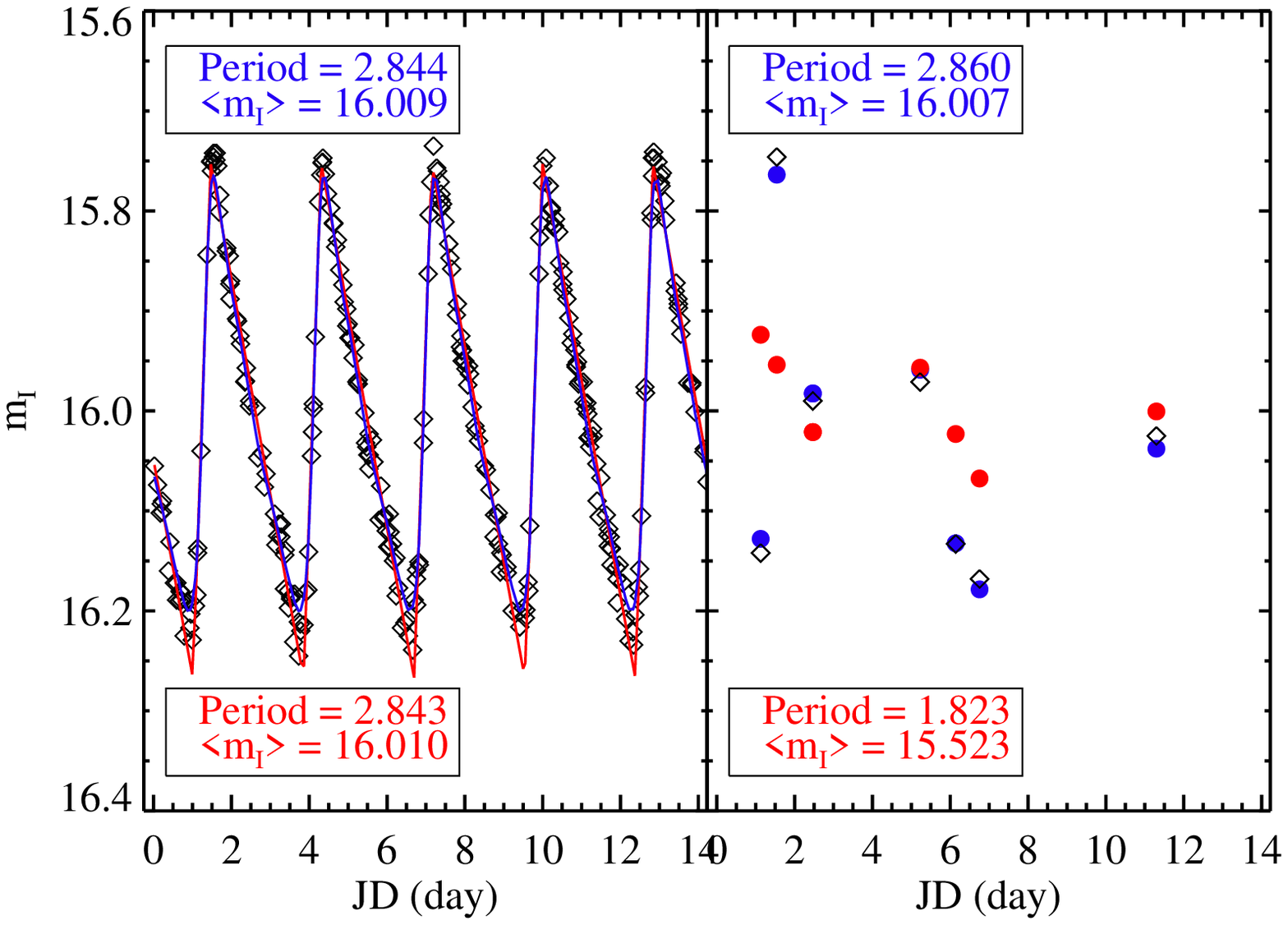} \\
\plottwo{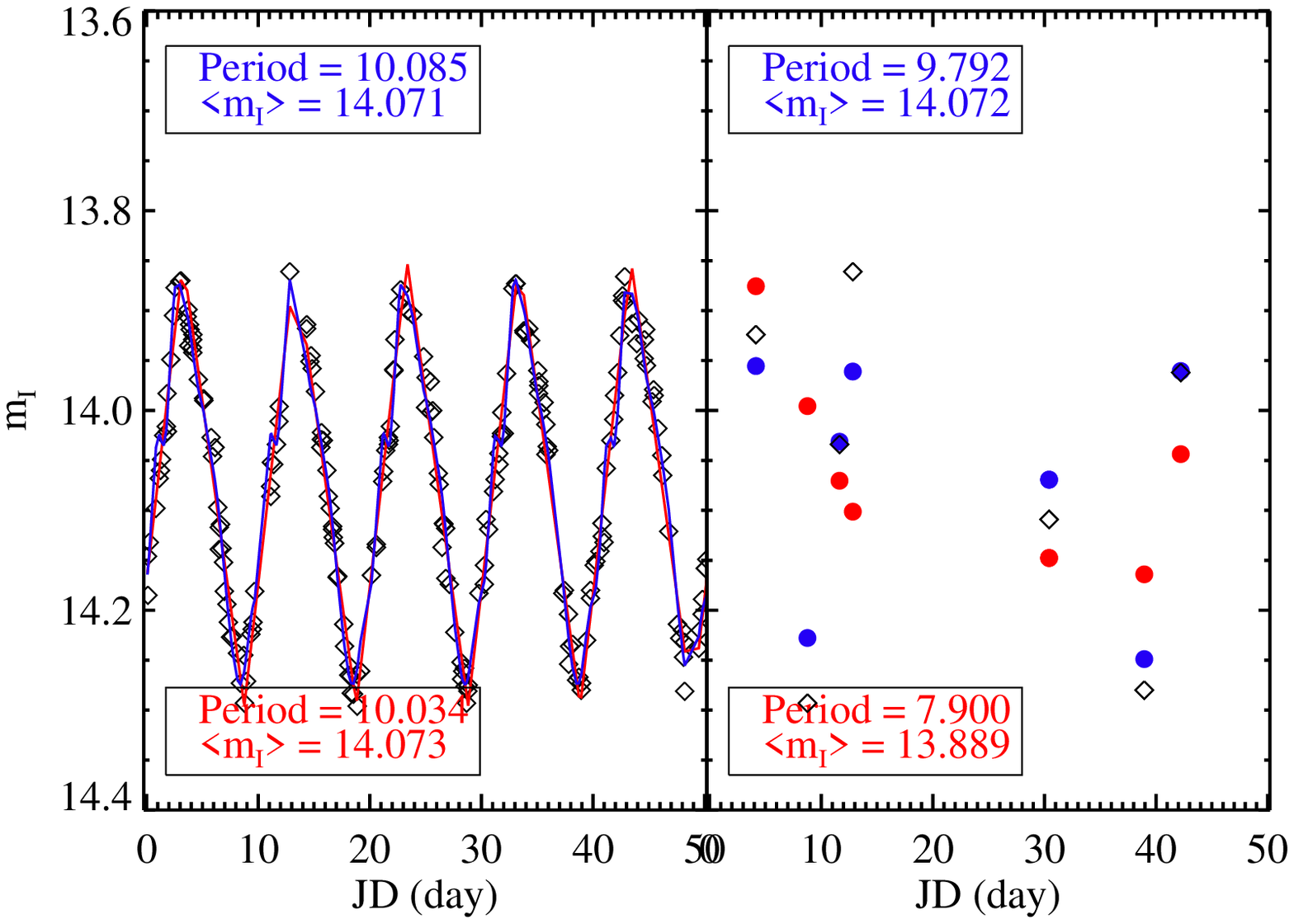}{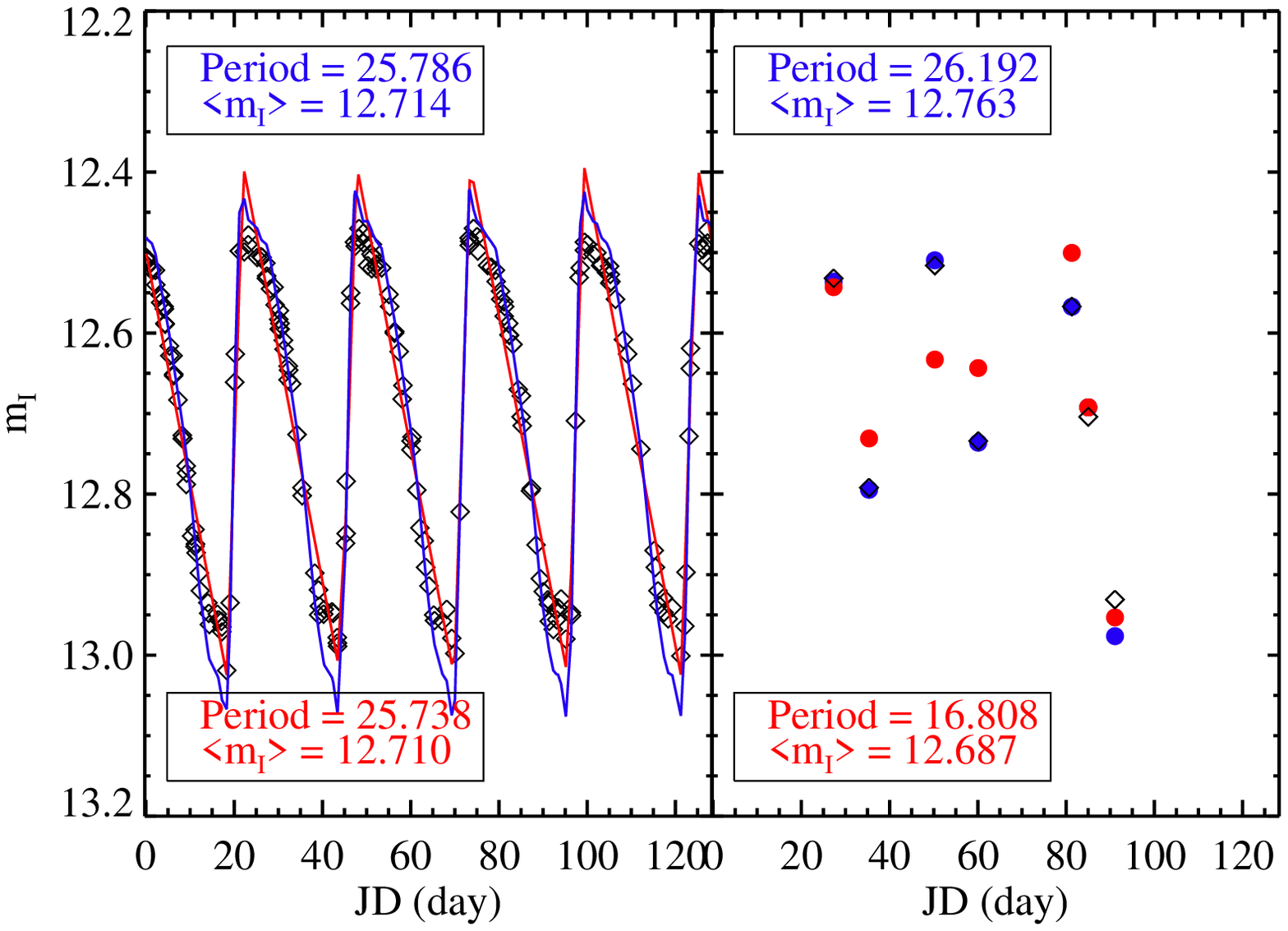}
\caption{ An example of how effective our templates can be at fitting sparsely sampled data.  In the left hand panels are the $I$-band light curves from Figure~\ref{example2} fitted with our LMC templates (blue) and a simple asymmetric saw-tooth function (red).  The right hand panel shows the results when 7 random points from the light curve are sampled and the the fitting is repeated.  In four of the 5 cases, the templates are able to fit accurate average magnitudes and periods.  The larger number of free parameters in the saw-tooth function prevents it from properly converging in the sparsely sampled cases.\label{saw_tooth}}
\end{figure*}

To illustrate how well our fitting procedure works, we compared our template fits to fits using a simple asymmetric saw-tooth function.  The results are plotted in Figure~\ref{saw_tooth}.  When the light curves are well sampled, the templates and saw-tooth converge to practically identical values.  In the sparsely sampled case (only 7 observational epochs), the templates return accurate fits in 4 out of 5 cases while the saw-tooth function fits fail in every case.  

In summary, our tests indicate that our PCA template technique can fit periods with an precision of $\pm0.1-0.3$ days with only 6 epochs and photometric precision of 0.01 mag.  If the number of epochs increases to 15 days, our uncertainty drops to $\pm0.03-0.2$ days.

\section{Converting Fits to Distances}\label{dist}

Having established that our templates can accurately fit Cepheid
periods and average magnitudes, we now point out some of the potential
pitfalls in using fitted parameters to derive accurate distances.  In
theory, distances from single Cepheids can be averaged together to
find a precise distance to a galaxy.  Besides the usual systematic
errors associated with photometry, there are additional caveats that
apply when observing a sample of Cepheids: (1) Cepheids used for a
distance calculation must be above the completeness limit of the
observations, otherwise, faint short-period stars will be
under-sampled and the final calculated distance will be biased
\citep{Sandage88,Freedman01}; (2) If the Cepheids do not sample the
full range of the instability strip, they can be offset from a
standard PL-relation.  \citet{Mager08} discuss how the scatter in the
PL-relation in the outskirts of NGC 4258 \citep{Macri06} is greatly
reduced because the Cepheids populate a limited region of the
instability strip; (3) Finally, there is always the risk that a
Cepheid may not be de-blended from a nearby optical/physical
companion.  Blending is expected to bias Cepheid distance measurements
to smaller values \citep{Stanek99,Mochejska00, Mochejska04}.

%In this case the light curve will look like that of a
%Cepheid, but the fitted magnitudes will be artificially bright

%The basic equations governing the calculation of a distance modulus are given by the PL-relation, the definition of the distance modulus, and reddening corrected distance modulus:

%\begin{eqnarray}
%M_{V,I}=a_{V,I}\log P+b_{V,I} \label{eq_pl}\\
%\mu_{V,I}=m_{V,I}-M_{V,I} \label{eq_dmod} \\
%\mu^0=\frac{ {\cal R}_I}{{\cal R}_I-{\cal R}_V}\mu_V -
% \frac{{\cal R}_V}{{\cal R}_I-{\cal R}_V}\mu_I  \label{eq_dust}
%\end{eqnarray}
%where ${\cal R}_V$ and ${\cal R}_I$ are the absorption-to-reddening ratios for the $V$ and $I$ filters and $\mu^0$ is the extinction corrected distance modulus.

\begin{figure*}
%\epsscale{.8}
\centering
\plottwo{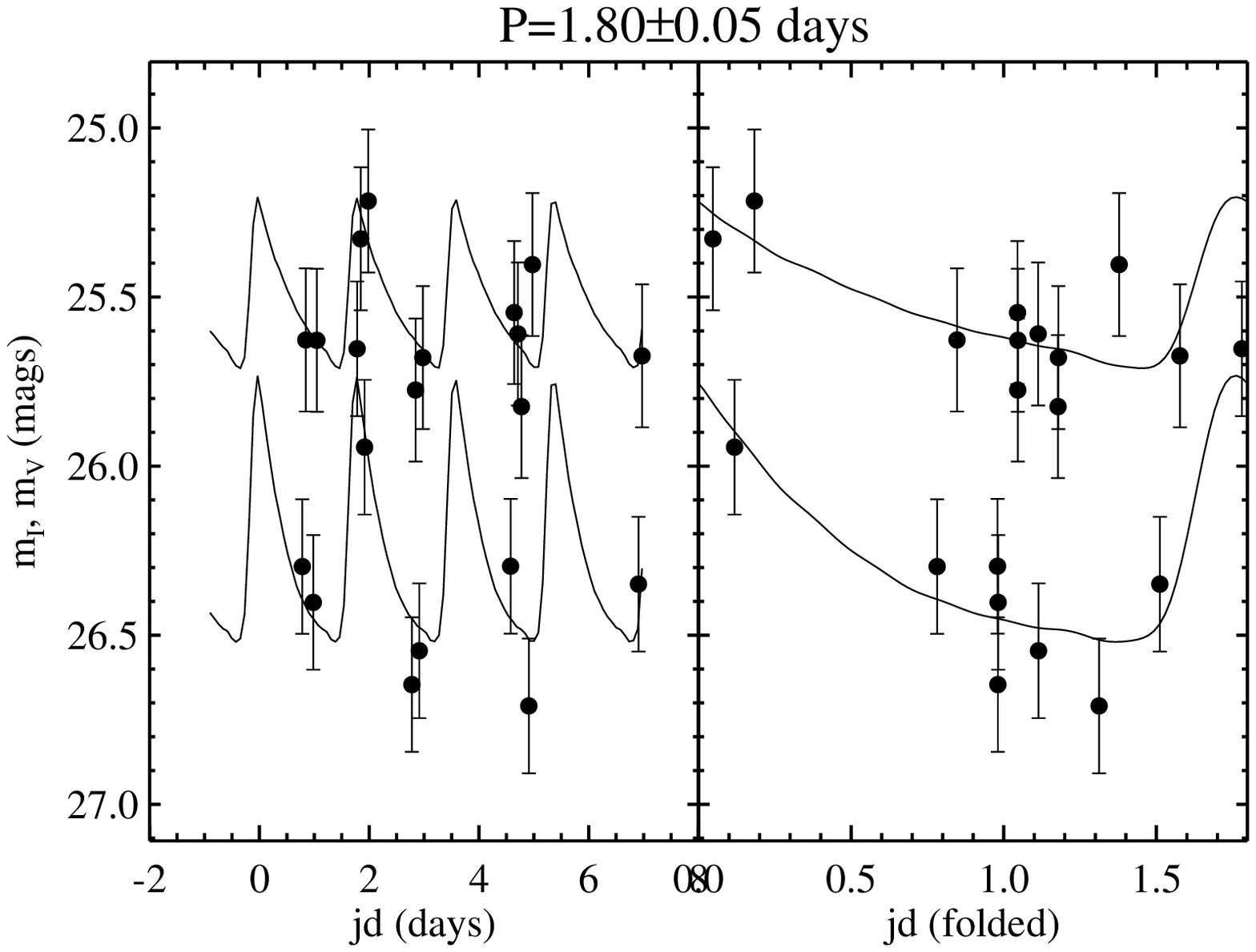}{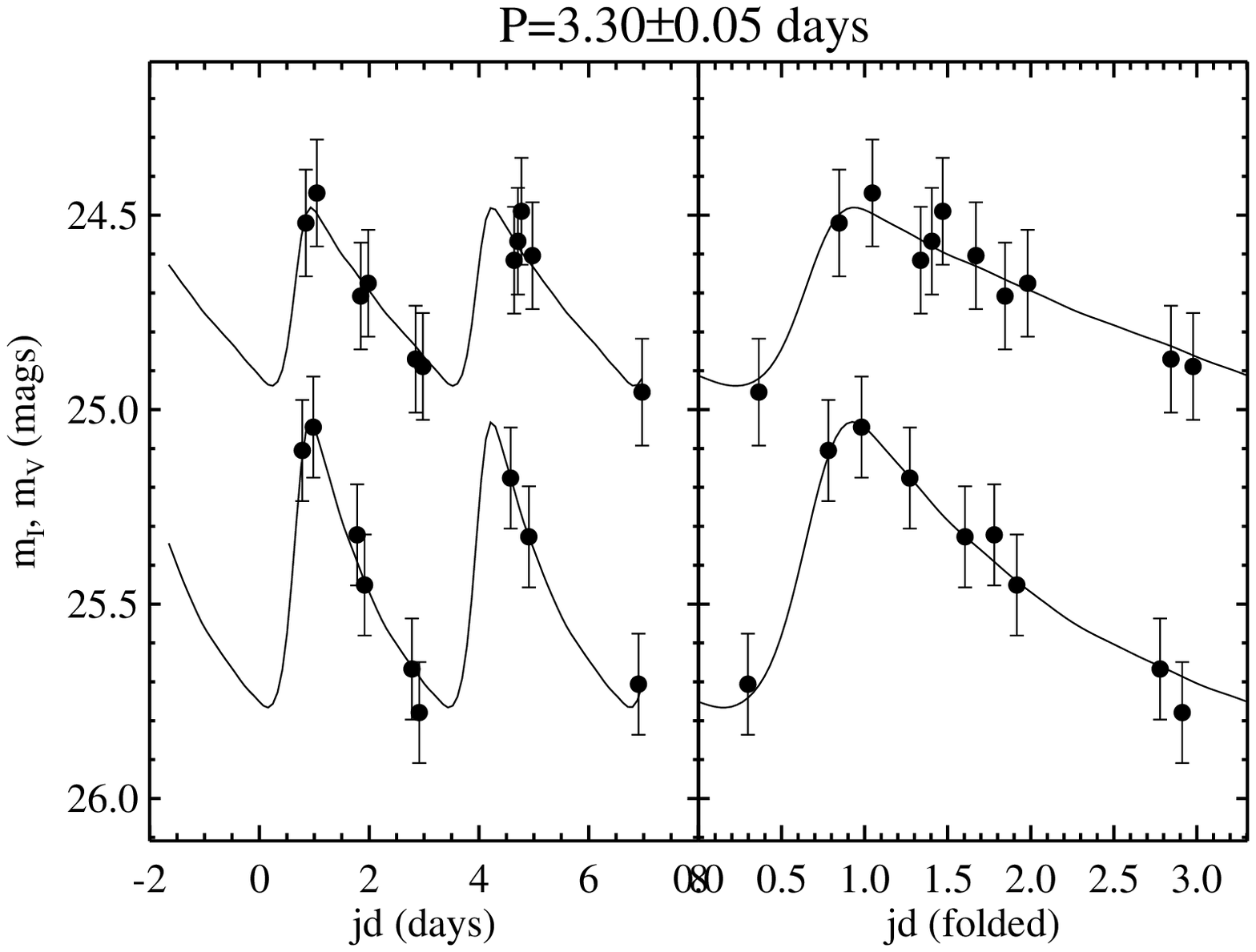}
\plottwo{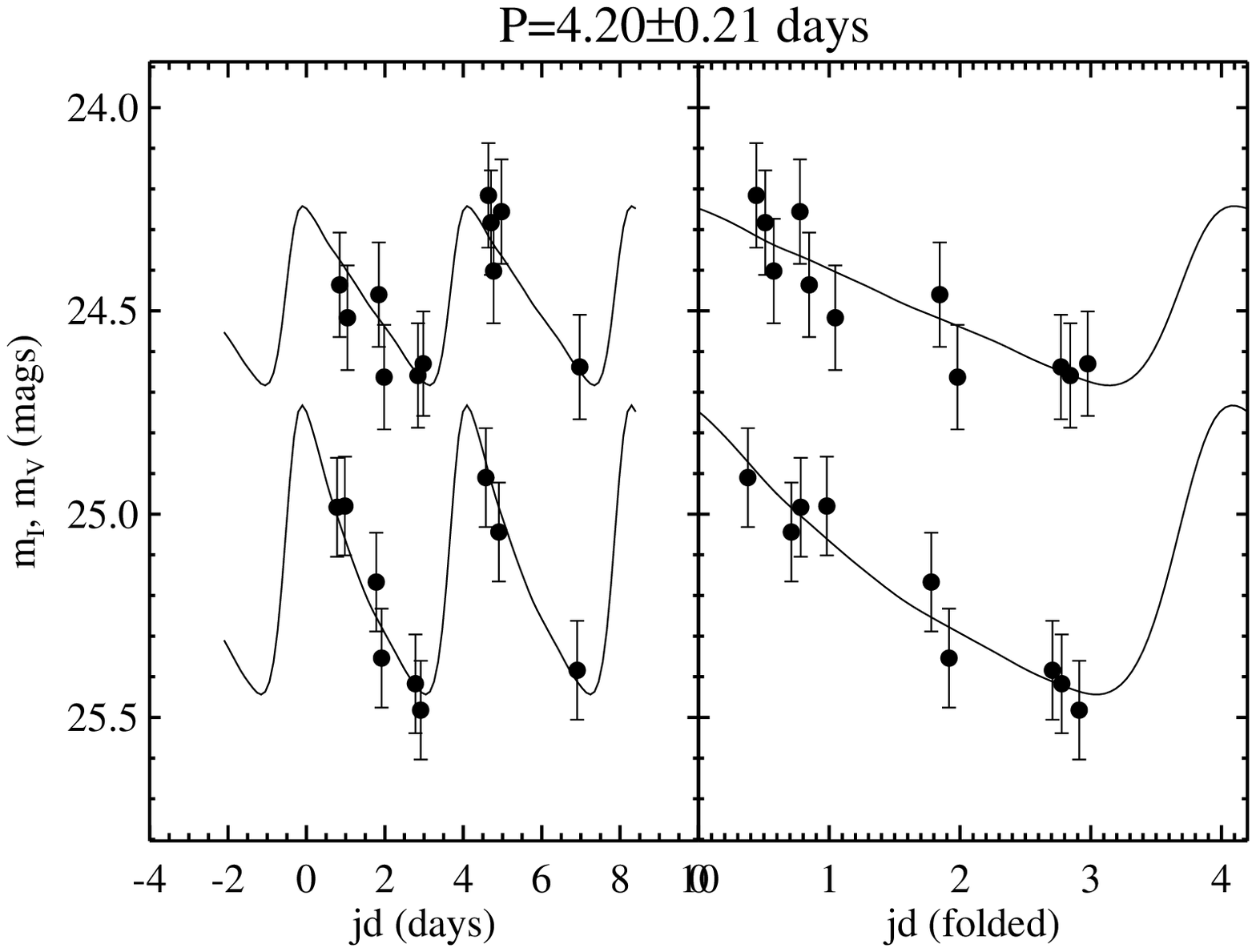}{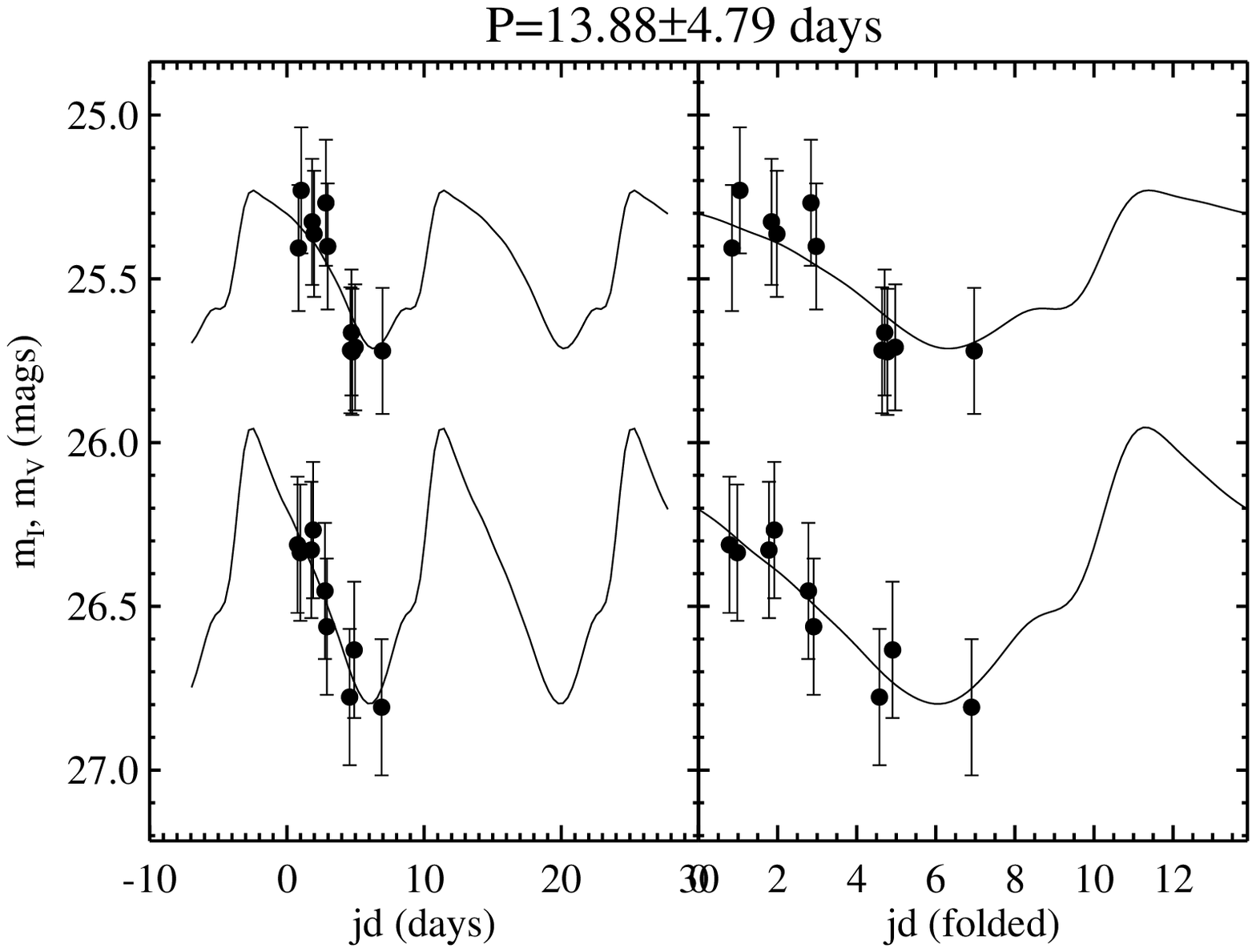}
\caption{Example of $V$ and $I$ light curves of Cepheids in M81 observed with HST and fit with our templates.  See \citet{McCommas08} for a larger sample of Cepheids and a full distance calculation to M81.  The fitted period in the lower right is poorly constrained due to the sparse phase coverage. \label{ex_lc}}
\end{figure*}

In a companion paper \citep{McCommas08}, we have successfully applied
our templates to HST data.  Figure~\ref{ex_lc} shows some examples of
how well the templates can fit noisy and sub-optimally sampled data.
The templates only start to fail for the star with the longest period,
where only half of the full phase is observed.

\subsection{Metallicity Effects on Light Curve Shape}\label{met_sec}

There is some question as to whether the shape of the light curve can
be used to determine the metallicity of a \ce
\citep{Paczy00,Kanbur02}.  Looking at the long period Cepheids
(Figure~\ref{pcas}), the first and second principal components have
systematically smaller values for the SMC stars compared to the LMC
and MW, suggesting the shapes might be intrinsically different.  We
use a 2D Kolmogorov-Smirnov test to compare the distributions in the
PCA-Period distributions (the MW, LMC, and SMC long period variables
all have very similar period distributions).  The LMC and MW are
consistent (Probability $>$ 0.12) with being drawn from the same
populations for all of the PCA vectors (i.e., the shape of the light
curve at a given period does not show significant differences).  The
SMC shows a significant difference between the MW (P=0.008) and LMC
(P=0.04) in the distribution of the first PCA vector, but not the
higher order PCA vectors.  While the SMC light curve shape is
different on average, there is little hope of assigning a metallicity
to an individual star based on its light curve shape, since the
intrinsic scatter within the SMC is larger than the differences from
the LMC or MW.  Even with the large number of stars in our sample, it
is difficult to differentiate between the low metallicity SMC Cepheids
and the higher metallicity LMC and MW Cepheids.  It is possible, in
theory, to observe enough long period stars that one could distinguish
if a population was high or low metallicity based on the PCA
distribution.  However, it would require a prohibitively large sample
size, as well as sufficient phase coverage that the shape parameters
could be measured robustly.

Unlike the long period variables, the short period Cepheids in the
three systems have very different period distributions.  However, the
PCA distributions overlap to a sufficiently large degree
(Figure~\ref{pcas}) that light curve shape cannot readily be used to
measure the metallicity of individual stars.

While we find metallicity does not alter light curve shapes
significantly, there is evidence that metallicity differences can
alter the PL-relation.  Several studies claim to find a metallicity
dependence \citep{Kennicutt98,Sakai04,Macri06,Saha06,Romaniello08,
Sandage08}, while others find that the PL-relation is constant
across galaxies \citep{Udalski01,Gieren05}.

\section{Conclusions}

We have extended the principal component analysis techniques of
\citet{Tanvir05} to generate template light curves of Cepheid
variables and first-overtone variables.  We have used a Monte Carlo
simulation to demonstrate how robustly our templates can be used to
fit Cepheid periods and magnitudes.  Unlike previous studies, we do
not limit ourselves to stars with periods longer than 10 days.
Finally, we demonstrate the effectiveness of our templates on HST data
and show that our techniques can be used to fit accurate periods and
luminosities even when the observations have not been optimally spaced
for observing variable stars.  Our templates open up a new regime of
distance measurement possibilities by enabling accurate fits for long
period, short period, and overtone Cepheids from noisy and sparsely
sampled observations.

%--------------------------------------
\acknowledgments 

We thank Andy Becker for helpful conversations.  Thanks to Anil Seth
and Greg Stinson for demanding more creative paper titles.  PY was
supported by the Harlen J. Smith Postdoctoral Fellowship.  JJD and PY
were partially supported through NSF grant CAREER AST-0238683 and the
Alfred P.\ Sloan Foundation. Support for this work was provided by
NASA through grant G0-10915 and AR-10945 from the Space Telescope
Institute, which is operated by the Association of Universities for
Research in Astronomy, incorporated under NASA contract NAS5-26555.
This work made use of the McMaster Cepheid Database, maintained by
Doug Welch.  This work made use of Craig Markwardt's totally awesome
IDL curve fitting code.

%\bibliography{../../../Papers/Bib_files/big_jabref}

\end{document}